\documentclass[journal,twocolumn]{IEEEtran}
\usepackage{cite}
\usepackage{amsmath,amssymb,amsfonts}
\usepackage{algorithmic}
\usepackage{graphicx}
\usepackage{textcomp}
\usepackage{wrapfig}
\usepackage{subfig} 
\usepackage{adjustbox}
\usepackage{multirow}
\usepackage{romannum}
\ifCLASSINFOpdf

\else

\fi

\begin{document}
	\title{Simultaneous estimation of wall and object parameters in TWR using deep neural network}
	
	\author{Fardin~Ghorbani,{} Hossein~Soleimani{}
		
		\thanks{Corresponding Author: Hossein Soleimani (email: hsoleimani@iust.ac.ir)}
		\thanks{F. Ghorbani, H. Soleimani, are with the School of Electrical
			Engineering, Iran University of Science and Technology, Tehran, Iran}
	}	
	
	\maketitle
	
	\begin{abstract}
		This paper presents a deep learning model for simultaneously estimating target and wall parameters in Through-the-Wall Radar. In this work, we consider two modes: single-target and two-targets. In both cases, we consider the permittivity and thickness for the wall, as well as the two-dimensional coordinates of the target's center and permittivity. This means that in the case of a single target, we estimate five values, whereas, in the case of two targets, we estimate eight values simultaneously, each of which represents the mentioned parameters. We discovered that when using deep neural networks to solve the target locating problem, giving the model more parameters of the problem increases the location accuracy. As a result, we included two wall parameters in the problem and discovered that the accuracy of target locating improves while the wall parameters are estimated. We were able to estimate the parameters of wall permittivity and thickness, as well as two-dimensional coordinates and permittivity of targets in single-target and two-target modes with 99\% accuracy by using a deep neural network model. 
	\end{abstract}
	
	
	\IEEEpeerreviewmaketitle
	
	\section{Introduction}

	\IEEEPARstart{R}{ecent} advancement in machine learning algorithms, particularly deep learning, and their penetration into other sciences have resulted in the solution of novel problems in various fields. The impact of machine learning on physics\cite{carleo2019machine}, signal processing\cite{ghorbani2020eegsig}, metasurface design\cite{jafar2020tco,ghorbani2021deep,ghorbani2021deep2}, antenna design\cite{sharma2020machine}, electromagnetic compatibility\cite{medico2018machine} and other fields has demonstrated the high flexibility and ability of these techniques to improve previous results and solve new problems in a variety of fields. These characteristics of machine learning, particularly deep learning, to discover hidden patterns in signals make it an excellent tool for analyzing radar signals. Among the applications of machine learning in radar, we focus on Through-the-Wall Radar (TWR) and investigate how it can solve new problems in this field that conventional methods cannot \cite{wood2020through,ghorbani2021deepwall}.
	
	In general, there are two types of methods for estimating wall and target parameters: conventional methods and machine-based methods. Methods such as time-delay\cite{protiva2011estimation}, filter-based methods \cite{jin2012image}, M-Sequence sensor, and continuous basis estimator\cite{fereidoony2017m} are conventional methods.
	Additionally, machine learning-based methods can be classified into two categories: those that utilize conventional machine learning algorithms such as SVM and methods based on deep learning.

	Zhang et al.\cite{zhang2016real} developed an SVM-based method for two-dimensional locating under a homogeneous wall and a circular metal cylinder object. They also attempted to estimate the wall parameters using the same method, which is based on SVM\cite{zhang2015application,zhang2016efficient}. In \cite{zhang2020robust}, a 3D positioning method is proposed for a
	homogeneous wall for a spherical metallic object
	using an extreme learning machine. Wood et al. \cite{wood2020through} investigated a machine learning (ML) approach for three objectives, one
	of which is predicting the location of targets. This work performed two-dimensional positioning with a circular object sing the K Nearest Neighbors (KNN) algorithm and a homogeneous non-magnetic wall. Common methods used in TWR for locating targets, estimating wall parameters such as permittivity, wall thickness, etc., separately. In some of these methods, the wall effect must be removed to locate targets because ambiguities in the wall parameters distort the imaging and shift the target location. On the other hand, conventional methods are incapable of estimating the target characteristics and instead concentrate exclusively on the target location. In this work, we presented a model for simultaneously estimating the wall permittivity and thickness, as well as the two-dimensional location of targets and type of targets using a deep learning approach. In \cite{ghorbani2021deepwall}, we proposed two-dimensional positioning for the case that the wall is modeled as a complex electromagnetic wall by presenting three deep learning models. Now, assuming the wall is a perfect non-magnetic dielectric, we attempt to estimate the wall parameters, such as permittivity and thickness, as well as the target parameters, such as the two-dimensional location of the targets and the target permittivity, simultaneously using a deep neural network.

	\section{METHODOLOGIES}
	
	\subsection{Deep Learning}
	Deep learning is a subset of machine learning and artificial intelligence that closely mimics the process by which the human mind acquires knowledge. This type of learning is critical in data science, which also encompasses statistics and forecasting modeling. In-depth learning is extremely beneficial for data scientists who are responsible for collecting, analyzing, and interpreting large amounts of data, as it speeds up and simplifies the process. Deep learning is actually the process of learning through neural networks with numerous hidden layers, and the deeper these layers go, the more complex and complete the models become. Deep learning is distinguished by its approach to solving problems. Working with conventional machine learning algorithms such as SVM, etc., begins with manual feature extraction. Then, a machine learning model is constructed using these features. However, the process of deep learning is designed so that the computer automatically detects and extracts the relevant features. Additionally, deep learning employs end-to-end learning, in which raw data is fed into the neural network and assigned a task, such as categorization. The deep learning model automatically learns how to do this. Deep learning models are taught using large sets of labeled data and neural network architectures that automatically learn features from the data without the need to extract them manually. 
	A neural network is composed of multiple layers of neurons. Typically, neural networks consist of three layers: input, hidden, and output. As the number of layers and neurons in each hidden layer increases, the model becomes more complex. As the number of hidden layers and neurons in our network increases, it transforms into a deep neural network, referred to as deep learning.
	
	\begin{figure}[h]
		\centering
		\includegraphics[scale=0.08]{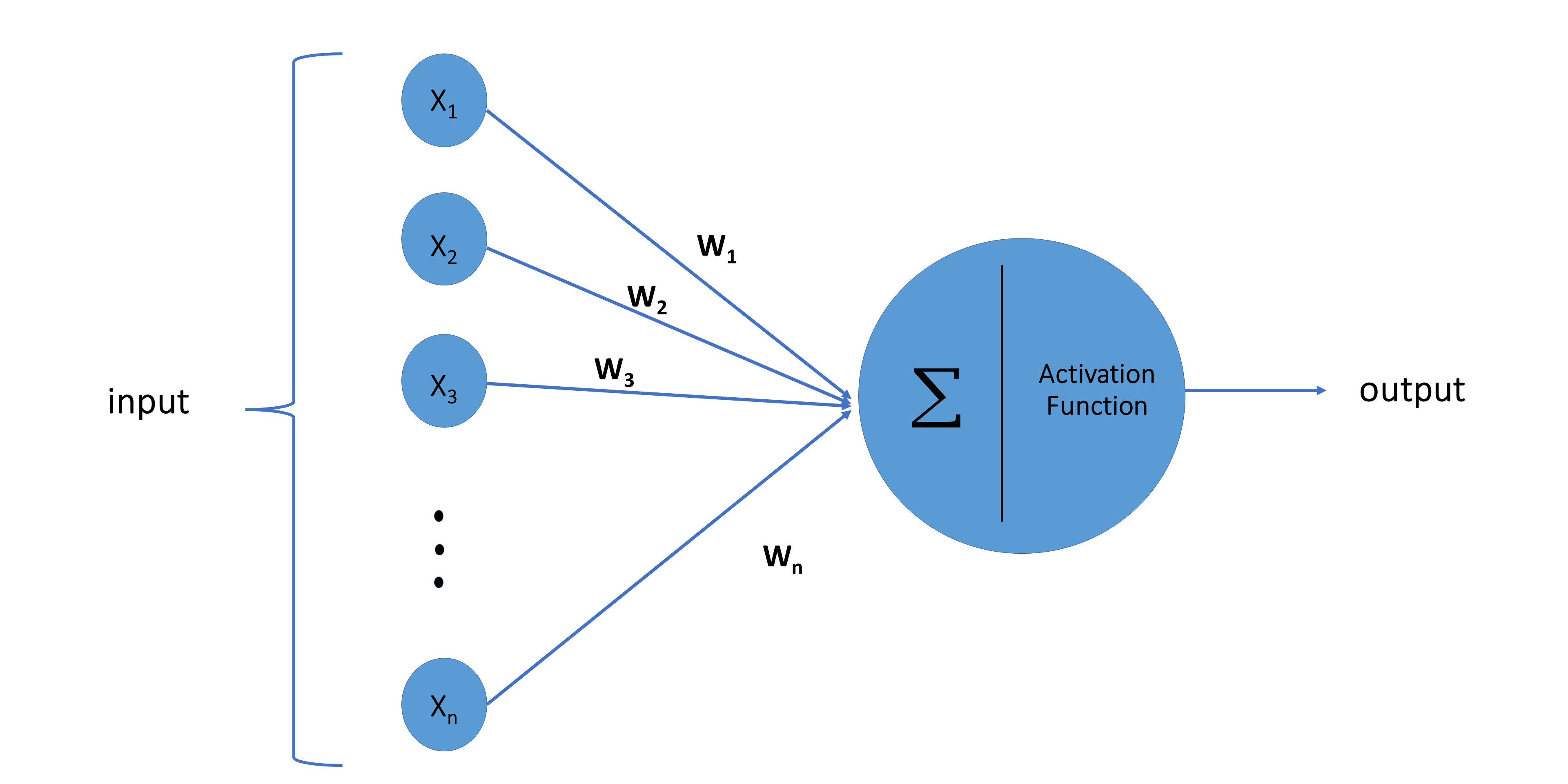}
		\caption{An overview of an artificial neuron} 
	\end{figure}
	
	Figure 1 shows an overview of artificial neurons. The inputs (input neurons) are $X_{1}$, $X_{2}$, ... . Each $X_{i}$ in a neural network has a weight, denoted by $W_{i}$. Indeed, each input is weighted independently. The neural network sum function (sigma) then adds the products of the X and W, and the activation function calculates the output value based on this calculation. The output of a neuron can be expressed as if the activation function is represented by $ F (u) $ and b is the bias value. The neuron output  can be described as follows:
	\begin{equation}
	Y =F(\sum\limits_{i=1}^n W_{i}X_{i}+b_{i})
	\end{equation}

	\subsection{FDTD}
	The finite difference time-domain (FDTD) is a method for solving Maxwell's equations. The equations of Ampere's Law and Faraday's Law can be written as follows:
	
	\begin{equation}
	\begin{aligned}
	\quad\nabla\times{H} = \frac{\partial{D}}{\partial t} = \varepsilon \frac{\partial{E}}{\partial t} \quad & ,   & \quad \text{(Faraday's Law)} \\[5pt]
	\quad\nabla\times{E} = -\frac{\partial{B}}{\partial t} = \mu \frac{\partial{H}}{\partial t} \quad & ,  & \quad \text{(Ampere's Law)}   \\[5pt]
	\end{aligned}
	\end{equation}
	
	We use  $TM_{z} $ polarization to rewrite the Ampere's Law and Faraday's Law as \cite{schneider2010understanding}:
	
	\begin{equation}
	\epsilon\frac{\partial{E}}{\partial t} =\quad\nabla\times{H} = \begin{vmatrix}
	\hat{a_{x}} & \hat{a_{y}} & \hat{a_{z}} \\ 
	\frac{\partial}{\partial x} & \frac{\partial}{\partial y} & 0\\ 
	H_{x} & H_{y} & 0\\ 
	\end{vmatrix} = \hat{a_{z}}(\frac{\partial H_{y}}{\partial x}-\frac{\partial H_{x}}{\partial y})
	\end{equation}
	
	\begin{equation}
	-\mu\frac{\partial{H}}{\partial t} =\quad\nabla\times{E} = \begin{vmatrix}
	\hat{a_{x}} & \hat{a_{y}} & \hat{a_{z}} \\ 
	\frac{\partial}{\partial x} & \frac{\partial}{\partial y} & 0\\ 
	0 & 0 & E_{z}\\ 
	\end{vmatrix} = \hat{a_{x}}\frac{\partial E_{z}}{\partial y}-\hat{a_{y}}\frac{\partial E_{z}}{\partial x}
	\end{equation}
	
	The scalar equations for $TM_{z} $ are obtained from (3) and (4):
	
	\begin{equation}
	-\mu\frac{\partial{H_{x}}}{\partial t} = \frac{\partial E_{z}}{\partial y}
	\end{equation}

	\begin{equation}
	\mu\frac{\partial{H_{y}}}{\partial t} = \frac{\partial E_{z}}{\partial x}
	\end{equation}
	
	\begin{equation}
	\epsilon\frac{\partial{E_{z}}}{\partial t} = \frac{\partial H_{y}}{\partial x}-\frac{\partial H_{x}}{\partial y}
	\end{equation}

	Equations (5)-(7) can be written in finite-differences form, and future fields can be expressed in terms of past fields due to the space-time discretization. The indexes $m$ and $n$ denote the spatial steps in the x and y directions, respectively, and the index q corresponds to the temporal step. Additionally, the spatial step sizes are $\Delta x$ and $\Delta y$ in the x and y directions, respectively.
	The finite difference approximation of (5) expanded about
	the space-time point (m$\Delta x$, (n + 1/2)$\Delta y$, q$\Delta t$). The resulting equation is:
	
	\begin{align}
	\begin{split}
	-\mu\frac{H_x^{q+\frac{1}{2}}[m,n+\frac{1}{2}] - H_x^{q-\frac{1}{2}}[m,n+\frac{1}{2}]}{\Delta t}=\\
	\frac{E_z^{q}[m,n+1] - E_z^{q}[m,n]}{\Delta y}
	\end{split}
	\end{align}
	
	The equation can be rewritten as follows for future value in terms of past value:
	
	\begin{align}
	\begin{split}
	H_x^{q+\frac{1}{2}}[m,n+\frac{1}{2}] = H_x^{q-\frac{1}{2}}[m,n+\frac{1}{2}]-\\ \frac{\Delta t}{\mu \Delta y} (E_z^{q}[m,n+1] - E_z^{q}[m,n])
	\end{split}
	\end{align}
	
	We can also write for Equation (6) and (7) expanded about
	the space-time point ((m + 1/2)$\Delta x$, n$\Delta y$, q$\Delta t$) and (m$\Delta x$, n$\Delta y$, (q + 1/2)$\Delta t$), respectively:
	
	\begin{align}
	\begin{split}
	H_y^{q+\frac{1}{2}}[m+\frac{1}{2},n] = H_y^{q-\frac{1}{2}}[m+\frac{1}{2},n]+\\ \frac{\Delta t}{\mu \Delta x} (E_z^{q}[m+1,n] - E_z^{q}[m,n])
	\end{split}
	\end{align}
	
	\begin{align}
	\begin{split}
	E_z^{q+1}[m,n] = E_z^{q}[m,n] + \frac{\Delta t}{\epsilon \Delta x} \lbrace H_y^{q+\frac{1}{2}}[m+\frac{1}{2},n] - \\
	H_y^{q-\frac{1}{2}}[m-\frac{1}{2},n] \rbrace - \frac{\Delta t}{\epsilon \Delta y}v\lbrace H_y^{q+\frac{1}{2}}[m,n+\frac{1}{2}]\\ -H_y^{q+\frac{1}{2}}[m,n-\frac{1}{2}]  \rbrace
	\end{split}
	\end{align}

	\begin{figure*}[th]
		\centering
		\includegraphics[scale=0.17]{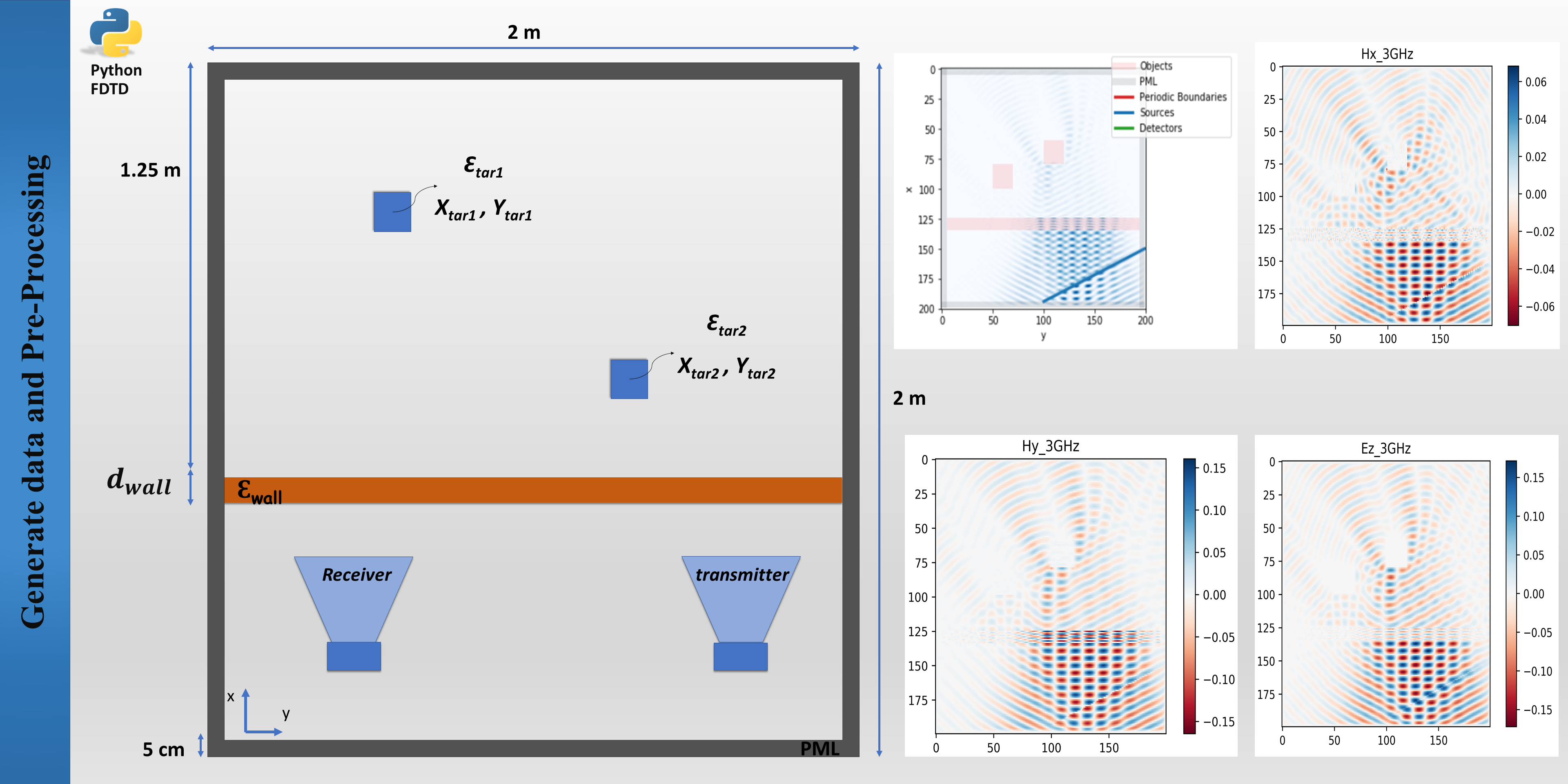}
		\caption{Environment and parameters required for generating the dataset. Simulations performed for the fields $ E_ {z}, H_ {x}, H_ {y} $ are also available in the figure.} 
	\end{figure*}

	\subsection{Data Gathering}
	We consider a two-dimensional TWR problem and conduct simulations using Python's FDTD library \cite{fdtdpythonlib}.
	We consider a 30cm square with a $\varepsilon_{r} = 80$ as the target. We generate a plane wave with a frequency of 3 GHz using a line source in the FDTD library. We use $TM_{z} $ polarization, which implies that $E_{x} = E_{y} = H_{z} = 0$ and $E_{z} , H_{x} , H_{y} $  are non-zero. The source's emitted wave hits the wall, and some of it returns, while the remainder passes through the wall, hits the target, and scatters away from the target. Finally, the scattered wave is received by the detector. We extract the fields of $E_{z} , H_{x} , H_{y} $ and use them to create the required dataset.
	We assume a homogeneous wall  with  $\mu = 1$, $\sigma = 0$, and  $\varepsilon_{r}$ used for data generation varies from 3 to 9.
	To create a single-target dataset, we move the target in the two-dimensional space specified in Figure 1, change the target permittivity from 5 to 85, the wall permittivity from 3 to 9, and the wall thickness from 10 to 20 cm. As a result, we generate 16,000 datasets in this mode. The two-targets mode is similar, except that instead of one object, we have two, introducing three additional parameters to the problem, including the target's two-dimensional location and permittivity. In this case, we generated 58,000 data sets.
	We allocated $70\%$ of the dataset for training, $15\%$ for the validation dataset, and $15\%$ for the test dataset.
	
	\begin{figure*}[th]
		\centering
		\includegraphics[scale=0.17]{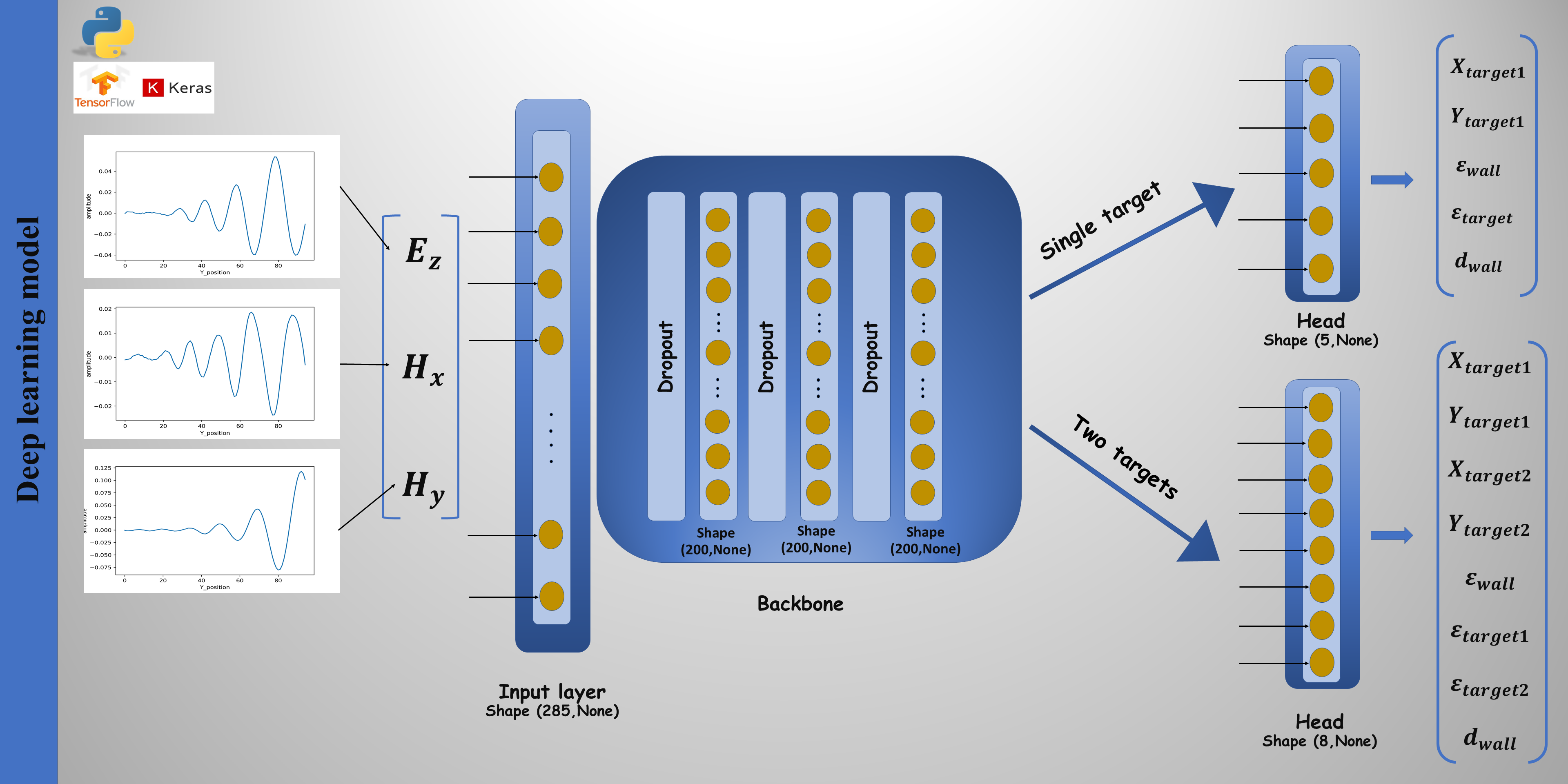}
		\caption{overview of the deep neural network architecture presented. The network inputs are $E_{z} , H_{x} , H_{y} $ fields and estimate the output according to one or two objectives or specified parameters for us.} 
	\end{figure*}
	
	\section{NUMERICAL AND EXPERIMENTAL RESULTS}
	We used a Deep Neural Network (DNN) to estimate both object and wall parameters concurrently in this work. Python is used to implement the DNN algorithm, as well as the Tensorflow and Keras frameworks \cite{chollet2015keras}. For this purpose, we presented a model in which the network input and network backbone, which are the hidden layers of the neural network, are the same for single-target and two-targets modes. However, the network output for these modes differs. In the single-target mode, we estimate five parameters: the target's two-dimensional location, wall permittivity, target permittivity, and wall thickness. As a result, we consider the number of neurons in the last layer to be five. In the case of two targets, we estimate three additional parameters: the 2D location and the permittivity of the second target.
	We used the Relu activation function for this neural network's first and middle layers but the linear activation function for the final layer. Equations 12 and 13 illustrate the activation of the Relu and Linear functions.
	
	\begin{equation}
	f(x) = \begin{cases}
	0 & x \leq 0 \\
	x & x > 0
	\end{cases} 
	\end{equation}

	\begin{equation}
	f(x) = x
	\end{equation}

	We trained the network using a batch size of 20, and a learning rate of 0.001, as well as the Adam optimizer and the Mean Squared Logarithmic Error (MSLE) loss function, as defined below:
	
	\begin{align}
	L(y,\hat{y}) = \frac{1}{N} \displaystyle\sum\limits_{i=0}^N (\log(y_{i}+1)-\log(\hat{y_{i}}+1))^2
	\end{align}
	
	y is the actual value, $\hat{y}$ is the estimated value, and $N$ is the total number of data. In fact, MSLE is the mean of the squared differences between the actual and estimated values after log transformation. 
	We sequentially combined the Dense and Dropout layers in order to achieve higher accuracy and less loss in the network's backbone.
	The network is trained for 200 epochs, and the Loss diagram for the train and validation datasets is shown in Figure 2 in both single and two target modes.
	Additionally, Table \Romannum{1} contains the accuracy and loss obtained on the validation and test datasets.

	\renewcommand{\arraystretch}{1.4}
	\begin{table}[]
		\centering
		\caption{The value of accuracy and loss for validation and test data.}
		\resizebox{1.3\textwidth}{!}{\begin{minipage}{\textwidth}
				\begin{tabular}{c|c|c|}
					\cline{2-3}
					& \textbf{single target} & \textbf{two targets} \\ \hline
					\multicolumn{1}{|c|}{\textbf{validation loss}}     & 0.052                  & 0.065                \\ \hline
					\multicolumn{1}{|c|}{\textbf{test loss}}           & 0.055                  & 0.063                \\ \hline
					\multicolumn{1}{|c|}{\textbf{validation accuracy}} & $\%99$                     & $\%98$                    \\ \hline
					\multicolumn{1}{|c|}{\textbf{test accuracy}}       & $\%99$                      & $\%99$                    \\ \hline
				\end{tabular}
		\end{minipage}}
	\end{table}

	\begin{figure}[t]
		\centering
		\subfloat[single target loss]{\includegraphics[width=5.7cm]{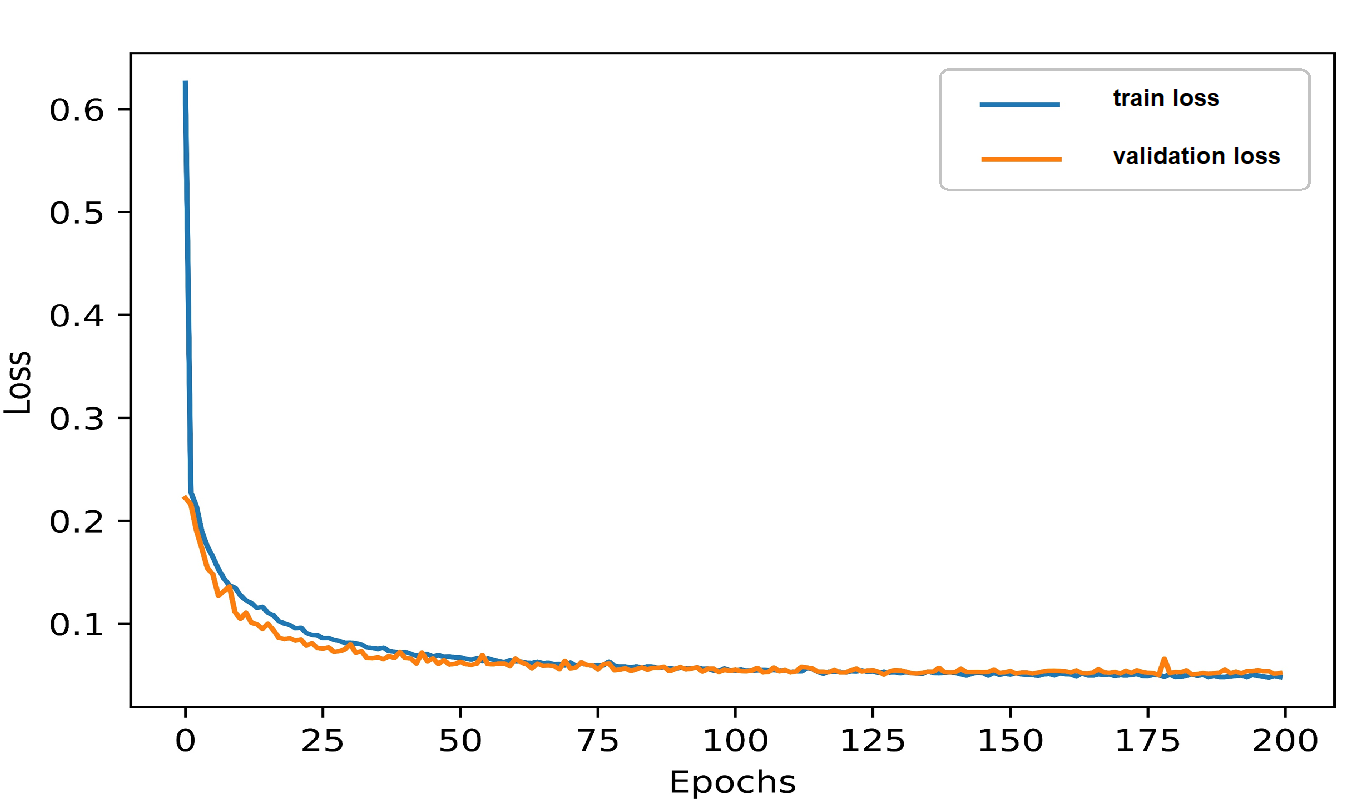}}
		\qquad
		\subfloat[two targets loss]{\includegraphics[width=5.7cm]{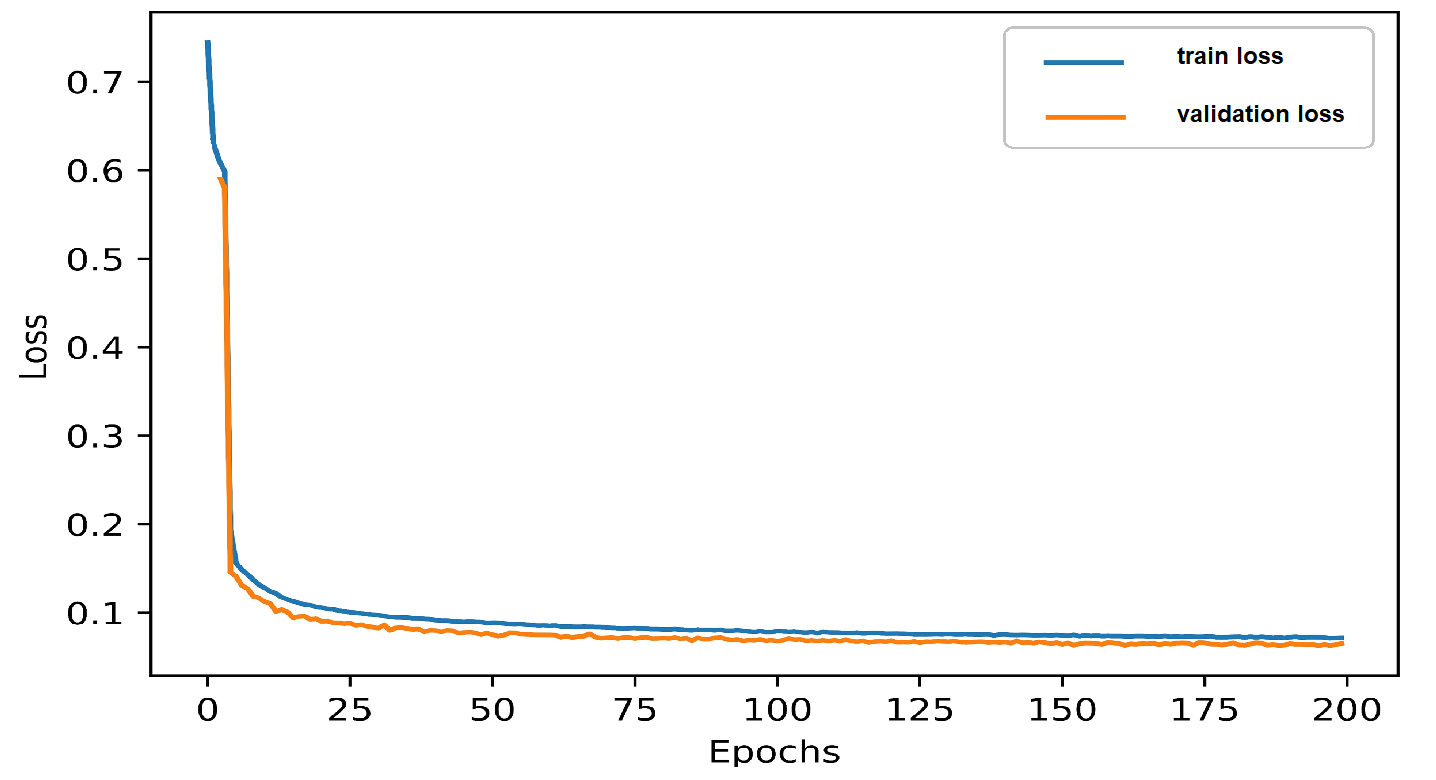}}
		\qquad
		\caption{loss diagrams in single and two targets locating.}
		\label{2fig}
	\end{figure}
	
	As it turns out, by including some target and wall specifications in the locating problem, we were able to improve the accuracy of the locating while also accurately estimating the wall and target material parameters. Indeed, each problem involves a large number of parameters. When we attempt to solve a problem using machine learning, if we enter all of the parameters involved in the problem, the model presented by us can better learn the relationship between the inputs and outputs, thereby increasing the accuracy of the solution. We observed that in the problem of two-dimensional positioning, it is sufficient to include additional parameters associated with the signal received by the receiver in order to achieve high positioning accuracy. By including the permittivity of the target and the wall, as well as the wall thickness, in the problem, we discovered that the proposed model not only improved positioning accuracy but also allowed for the integration of other critical parameters that had previously been estimated separately using a separate algorithm; all of these parameters were estimated using the same deep learning model that was used to locate the targets.

	\section{CONCLUSIONS AND DISCUSSION}
	We present a model for simultaneously estimating the target and wall parameters using a deep neural network. The target parameters include the two-dimensional location and permittivity of the targets, as well as the thickness and permittivity of the wall. We considered two modes, one with a single target and one with two targets, both of which required two parameters for the wall specification and three parameters for each target. In this work, we generated a dataset by varying the parameters involved in the problem and then presented a deep learning model that allows the mentioned parameters to be estimated for various targets by simply changing the model's end layer. By incorporating the parameters contained in the received signal into the receiver, we were able to improve the location accuracy while simultaneously estimating parameters such as wall thickness and target and wall permittivity.
	
	\appendices

	\ifCLASSOPTIONcaptionsoff
	\newpage
	\fi

	\bibliographystyle{IEEEtran}
	\bibliography{myref}
	
\end{document}